\documentclass{article}

\usepackage{amsfonts}
\usepackage[margin=1in]{geometry}
\usepackage[affil-it]{authblk}
\usepackage{graphicx}
\usepackage{epstopdf}
\usepackage{cite}
\usepackage{amsmath}
\usepackage{xcolor}
\usepackage{amssymb}
\usepackage{textcomp}

\newcommand{\comm}[1]{}

\newcommand{\SI}{\boldsymbol{\sigma}}

\newcommand{\EP}{\boldsymbol{\varepsilon}}
\newcommand{\VE}{\mathbf}

\newcommand{\Srr}{\sigma_{rr}}

\begin{document}
\title{Analysis of residual stresses in thermoelastic multilayer cylinders}

\author{Gennaro Vitucci%
  \thanks{Corresponding author. Room 501, Department of Mathematics, IMPACS, Aberystwyth University, Ceredigion, SY23~3BZ, UK. Email: \texttt{gev4@aber.ac.uk}}}
\author{Gennady Mishuris}
\affil{Department of Mathematics, IMPACS, Aberystwyth University, Ceredigion, SY23~3BZ, UK}

\date{}
\maketitle

\begin{abstract}
The topic faced here is the modeling of an axisymmetric multilayer structure. An exact analytical formulation is proposed in the framework of the plane strain problem for a cylinder encircled by annular layers. Isotropic linear thermoelastic materials constitute the body. Perfect and imperfect contacts between the layers are made available for the analysis. The derived exact solution is compared to finite element simulations. Numerical applications are shown in order to study the dependency of the residual stress distribution on the constituent material properties during a cooling process. The role of residual stresses in brittle materials, particularly ceramics, is discussed.
\end{abstract}

\section{Introduction}
\label{sec:intro}
Ceramics present important and unique features, such as chemical inertness, high-temperature capability, hardness, which make them particularly suited to advanced engineering usages. Simultaneously, a well observed weakness is their fragile behavior. Unexpected brittle fractures with uncontrolled crack propagation speed arise after a threshold energy. The failure energy may be caused either by mechanical external loads and prestresses or by thermal induced differential deformations facing a material characterized by fortuitous and often designed inclusions (e.g. for fiber-reinforced ceramics).

It has been shown widely in literature that the macro-mechanical behavior with respect to crack propagation can be radically modified assembling multiple layers with different thicknesses and different material properties. Moreover, the full-thickness toughness does not depend only on the individual features of the layers, but also on their interactions. A property to be addressed is specifically the capability of laminar ceramics to deflect cracks.

It can be obtained by facilitating the crack bifurcation, event which can occur only in the case that the crack enters a compressed zone (\cite{r1, r13}). Thus a crucial role is played by controlling compressive prestresses arising during the cooling process below the joining temperature (\cite{r7,r8}). From a technological point of view, the creation of such conditions is committed to the choice of materials with different thermal expansion coefficients (\cite{r2, r9, r12}) and/or phase transformation diagrams upon cooling (\cite{r10, r11}). A second artifice to alter crack paths is the introduction of weak interfaces between layers in order to lead the deformation energy to obstacle the crack growth along the thickness (\cite{r5}). Delamination gives place to multiple stages fracturing and avoids catastrophic damage.

Similar observations show that a possible improvement of ceramics resistance is possible if the production process is carefully controlled, also considering that usually the size of the layers is of an order of tens of micrometers. On the other hand, aleatory defects arise in every industrial product and if they show features directed in the sense of favoring a catastrophic fracture, both economical and safety losses can invalidate ceramic materials usage for innovative applications. Indeed a practical disadvantage of ceramics is that it is not possible to predict which sample is brittle without testing it to destruction (\cite{r3}).

Prince Rupert's Paradox is a suggestive phenomenon that resumes the importance of the appearance of unintentional prestresses. The tip of a glass rod is melted before falling into room temperature water. The created drop-shaped solid demonstrates a very singular behavior: while its bulbous head turns very tough, its thin tail not only results fragile, but, when a fracture starts from there, it propagated with an extremely high velocity, in a self-sustained destruction wave, through the whole structure causing instantaneous disintegration (\cite{r6}). The reason is clearly to be researched in the arising of high residual stresses. Specifically, stretched internal zones of the head presented an elevate elastic energy which is suddenly released when even a small perturbation occurs (\cite{r3}).

In order to characterize similar features of brittleness we solve here the general plane strain axisymmetric problem of deformation in thermoelasticity which involves a cylinder inscribed into a series of concentric rings. The simple analytical solution can be advantageously adopted for following the residual stresses which derive from the difference of the interfacing materials or from the provided prestresses. Qualitative observations can be addressed by examining different components distribution inside the multilayer structure. Particularly, a graded distribution with respect to thermal expansion coefficients is shown. The engineering concern is mainly about the possibility allowed by the design to give place to compressive zones. Holding on the same structure and simply modifying Young's modulus of one of the layers, the additional effect of the presence of a soft layer is studied.


\section{Mechanical Model} 
\label{sec:mechmod}
The equilibrium, constitutive and kinematic equations in linear thermoelasticity for an isotropic material are
\begin{gather}
\nabla\cdot\SI=0,\label{eq:1}\\
\SI=\mathbf{C}(\EP-\mathbf{I}\hat{\alpha}\Delta T),\\
\EP=\frac{1}{2}\left(\nabla \VE{u}+{\nabla\VE{u}}^T\right)\label{eq:3},
\end{gather}
where $\SI$, $\EP$ and $\VE{u}$ are the stress tensor, the strain tensor and the displacement vector; $\mathbf{C}$, $\mathbf{I}$ the stiffness and and the identity tensors; $\hat{\alpha}$ and $\Delta T$ the coefficient of thermal expansion and the difference between the current temperature and the strain reference one. If two materials $(1)$ and $(2)$ are in contact by means of a very thin interphase, the latter can be classified differently depending on the relationships between its elastic properties and the surrounding media and mathematically substituted by transmission conditions (\cite{avila1999asymptotic,benveniste1985effective,klarbring1998asymptotic,mishuris2004imperfect}). Sticking to the definitions given by Benveniste and Miloh in \cite{benveniste2001imperfect}, three cases are considered in the present work: perfect, \textit{spring-type} (soft) and \textit{membrane-type} (stiff) interfaces. 

Let us consider the linear elastic deformation problem of a long axisymmetric structure undergoing an axisymmetric thermal and mechanical loading, so that its behavior can be adequately described in as axisymmetric plane strain. The only coordinate along which variable quantities arise is the radial one, say $r$, being $\theta$ and $z$ respectively the angular and axial coordinates; $\lambda$, $E$ and $\nu$ the material Lam\'{e}'s first parameter, Young's modulus and Poisson ratio. Our scope is to generalize the problem of a cylinder encircled by an arbitrary number $n$ of concentric layers, each of them characterized by possibly different thermomechanical parameters. An assigned uniform load or displacement on the outer border is also considered.

\begin{figure}%
\centering
\includegraphics[scale=0.6]{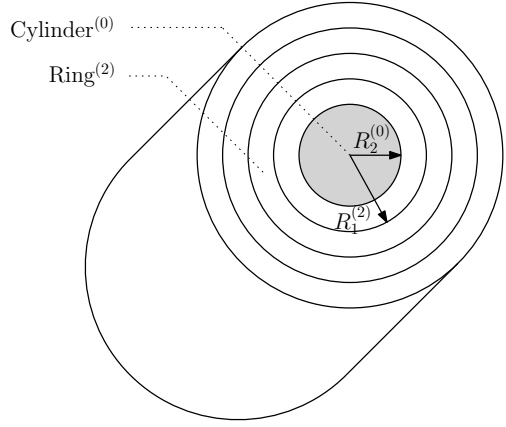}
\caption{Schematic undeformed configuration of a non-prestressed cylinder surrounded by multiple rings.}%
\label{fig:cyl}%
\end{figure}

The solution for each of the $n+1$ bodies ${(i)}$ is written by the fields Eq.\eqref{eq:Ufield} and \eqref{eq:Rfield}. If $\alpha^{(i)}=(1+\nu^{(i)})\hat{\alpha}^{(i)}$ is the coefficient of thermal expansion in plane strain for the $i$-th body (proportional to the classical coefficient $\hat{\alpha}$), then
\begin{align}
&u_{rr}^{(i)}=(a^{(i)}+\alpha^{(i)}\Delta T^{(i)})r+\frac{b^{(i)}}{r}, \label{eq:Ufield}\\
&\Srr^{(i)}=\frac{\lambda^{(i)}}{\nu^{(i)}}\left[a^{(i)}-(1-2\nu^{(i)})\frac{b^{(i)}}{r^2}\right],\qquad(i=0,1...,n).
\label{eq:Rfield}
\end{align}
The superscript $(0)$ labels the inner cylinder and $b^{(0)}=0$ descends from the symmetry condition. If the materials are not in mutual contact in their undeformed configuration they are assembled in such a way that the external radius of the $i$-th element is forced elastically to fit the internal radius of the $(i+1)$-th one -- namely \textit{prestressed}. It would be the case, for instance, of a cylinder which is inserted into a heated up ring before a homogeneous temperature is reached by the composite. The initial geometry of the multilayer body is completely described by the inner cylinder radius $R_2^{(0)}$ and by the internal and external radii $R_1^{(i)}$ and $R_2^{(i)}$ for each $i$-th ring ($i=1,2,...,n$), with the subscripts ${1}$ and ${2}$ meaning \textit{internal} and \textit{external} (see Fig.\ref{fig:cyl}).

In presence of soft contact, the equality of the radial displacements at the two sides of an interface is not guaranteed. Let $\beta^{(i)}$ be both the radial softness of the thin connecting layer ${(i)}$ between the components $(i)$ and $(i-1)$. \textit{Spring-type} transmission conditions lead to the boundary value problem:
\begin{align}
&u_2^{(i-1)}-u_1^{(i)}+{\sigma_1^{(i)}}{\beta^{(i)}}=R_1^{(i)}-R_2^{(i-1)}, \label{eq:D1D2}\\
&\sigma_2^{(i-1)}-\sigma_1^{(i)}=0, \label{eq:S1S2}
\end{align}
with $i=1,2,...,n$. From the latter equation on, where not specified, we refer simply to stresses $\sigma$ and displacements $u$ indicating radial ones, omitting subscripts for the coordinate system. If the $i$-th interface, instead, is of \textit{membrane type} with stiffness $\gamma^{(i)}$, the previous two equations are substituted via:
\begin{align}
&u_2^{(i-1)}-u_1^{(i)}=R_1^{(i)}-R_2^{(i-1)}, \label{eq:stD1D2}\\
&\sigma_2^{(i-1)}-\sigma_1^{(i)}+\gamma^{(i)}(u_1^{(i)}+R_1^{(i)})=0. \label{eq:stS1S2}
\end{align}
An ideal contact is taken into account simply setting $\beta^{(i)}$ and $\gamma^{(i)}$ to zero.
The outer boundary condition (BC) both in terms of axisymmetric uniform load $\overline{\sigma}$ ($\sigma$-BC) or radial uniform displacement $\overline{u}$ ($u$-BC) can be added by the equation:
\begin{equation}
\sigma_2^{(n)}=\overline{\sigma}
\quad\text{or}\quad
u_2^{(n)}=\overline{u} 
\label{eq:extBC}
\end{equation}
Being the $n+1$ unknowns $a^{(0)}$ and the $n$ couples $a^{(i)}$-$b^{(i)}$, the problem is completely stated by the set of $2n$ equations Eq.\eqref{eq:D1D2}-Eq.\eqref{eq:S1S2} or Eq.\eqref{eq:stD1D2}-Eq.\eqref{eq:stS1S2} plus Eq.\eqref{eq:extBC}.
For the sake of conciseness, it is useful to the define the following vectors and matrices ($i=1,2,...,n$):
\begin{equation}
\VE{M}_2^{(0)}=
\left[\begin{array}{c}
R_2^{(0)}\\ \frac{\lambda^{(0)}}{\nu^{(0)}}
\end{array}\right];\quad\quad
\VE{M}_2^{(i)}=
\left[\begin{array}{c c}
R_2^{(i)}&\frac{1}{R_2^{(i)}}\\
\frac{\lambda^{(i)}}{\nu^{(i)}}&-\frac{\lambda^{(i)}}{{R_2^{(i)}}^2}\frac{1-2\nu^{(i)}}{\nu^{(i)}}
\end{array}\right];
\end{equation}
\begin{equation}
\VE{M}_1^{(i)}=\left\{\begin{aligned}
&\text{soft interface:}&\quad&\left[\begin{array}{c c}
R_1^{(i)}-\frac{\beta^{(i)}\lambda^{(i)}}{\nu^{(i)}}&\frac{1}{R_1^{(i)}}+\frac{\beta^{(i)}\lambda^{(i)}}{{R_1^{(i)}}^2}\frac{1-2\nu^{(i)}}{\nu^{(i)}}\\
\frac{\lambda^{(i)}}{\nu^{(i)}}&-\frac{\lambda^{(i)}}{\nu^{(i)}}\frac{1-2\nu^{(i)}}{{R_1^{(i)}}^2}
\end{array}\right]\\
&\text{stiff interface:}&\quad&\left[\begin{array}{c c}
R_1^{(i)}&\frac{1}{R_1^{(i)}}\\
\frac{\lambda^{(i)}}{\nu^{(i)}}-\gamma^{(i)}R_1^{(i)}&-\frac{\lambda^{(i)}}{{R_1^{(i)}}^2}\frac{1-2\nu^{(i)}}{\nu^{(i)}}-\frac{\gamma^{(i)}}{R_1^{(i)}}
\end{array}\right]\\ 
\end{aligned}\right.\quad ;
\end{equation}
\begin{equation}
\VE{B}_2^{(n)}=\left\{\begin{aligned}
&\text{with $\sigma$-BC: }&\frac{\lambda^{(n)}}{\nu^{(n)}}&
\left[\begin{array}{c c}
1&-\frac{(1-2\nu^{(n)})}{{R_2^{(n)}}^2}\\
\end{array}\right]\\
&\text{with $u$-BC: }&&
\left[\begin{array}{c c}
1&\frac{1}{{R_2^{(n)}}}\\
\end{array}\right]
 \end{aligned}\right.\quad.
\end{equation}
The whole set of $2n$ conditions takes the form
\begin{equation}
\VE{M}\VE{a}=\VE{d},
\label{eq:system}
\end{equation}
where the following assembly holds:
\begin{equation}
\VE{M}=\small{\left[\begin{array}{c c c c c c c}
\VE{M}_2^{(0)}&-\VE{M}_1^{(1)}&&&&&\\
&\VE{M}_2^{(1)}&-\VE{M}_1^{(2)}&&&&\\
&&\ddots&\ddots&&&\\
&&&\VE{M}_2^{(i-1)}&-\VE{M}_1^{(i)}&&\\
&&&&\ddots&\ddots&\\
&&&&&\VE{M}_2^{(n-1)}&-\VE{M}_1^{(n)}\\
&&&&&&\VE{B}_2^{(n)}\\
\end{array}\right]};
\label{eq:matrsol}
\end{equation}
\begin{equation}
\VE{a}^T=\left[\begin{array}{c c c c c c c c c}
a^{(0)}&a^{(1)}&b^{(1)}&\cdots&a^{(i)}&b^{(i)}&\cdots&a^{(n)}&b^{(n)}\end{array}\right].
\end{equation}
The known terms are collected in the vector $\VE{d}$, whose components are
\begin{equation}
\left.\begin{aligned}
&d_{2i-1}=(1+\alpha^{(i)}\Delta T^{(i)})R_1^{(i)}-(1+\alpha^{(i-1)}\Delta T^{(i-1)})R_2^{(i-1)},\\
&d_{2i}=\left\{\begin{aligned}
&0&&\text{(soft interface)}\\
&-\gamma^{(i)}R_{1}^{(i)}(1+\alpha^{(i)}\Delta T^{(i)})&\quad&\text{(stiff interface)}
\end{aligned}\right. ,\\
&d_{2n+1}=\left\{\begin{aligned}
&\overline{\sigma}&\text{ for $\sigma$-BC}\\
&\overline{u}-\alpha^{(n)}\Delta T^{(n)}R_2^{(n)}&\text{ for $u$-BC}\\
\end{aligned}\right. ,
\end{aligned}\right. 
\end{equation}
with $i=1,2,...,n$.
From a computational point of view it is useful to underline that $\VE{M}$ is a bidiagonal block matrix. Consequently a sweep method (see \ref{sec:swm}, \cite{linkov1992difference,mishuris19972}) can be developed for its inversion, which takes into account that $\VE{B}_2{(n)}\in\mathbb{R}^{1\times 2} $ and $\VE{M}_2^{(0)}\in\mathbb{R}^{2\times 1}$ are not invertible.
It is clear that, once found the $2n+1$ constants contained in $\VE{a}$, they can be used for knowing displacements and stresses of every point of the domain.

If one needs to solve the similar problem of a hollow cylinder instead of a full one, it is sufficient ensure the first unknown $a^{(0)}$ to fulfill the internal boundary conditions in terms, for instance, of pressure or radial displacement constraints (see Eq.\eqref{eq:Ufield}-\eqref{eq:Rfield}). In that case the present solution enriches the work developed in \cite{vedeld2014stresses} in the framework of pipe and vessels design through the availability of prestresses and imperfect interfaces.

\section{Residual Stresses in Multilayer Ceramics}
\begin{figure}%
\centering	
\includegraphics[width=\textwidth]{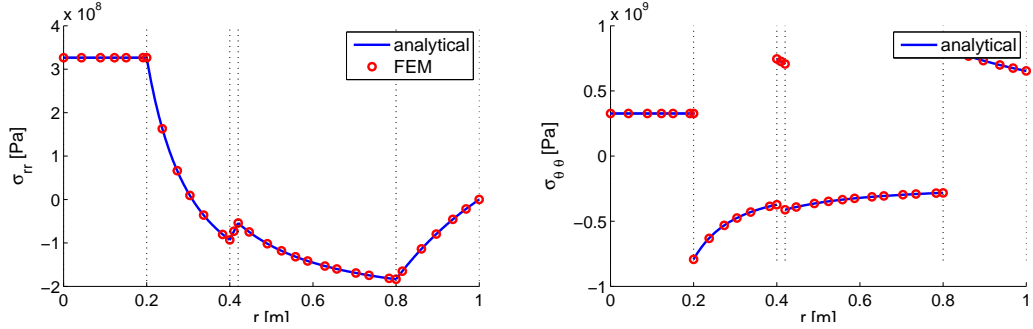}
\caption{Comparison of the residual $\sigma_{rr}$ and hoop $\sigma_{\theta \theta}$ stresses obtained for the case A by means of the analytical solution and a FE analysis run with COMSOL Multiphysics\textsuperscript{\circledR}. The mesh consists of 22710 triangular elements.}%
\label{fig:reshoop5}%
\end{figure}

\begin{figure}%
\centering	
\includegraphics[width=\textwidth]{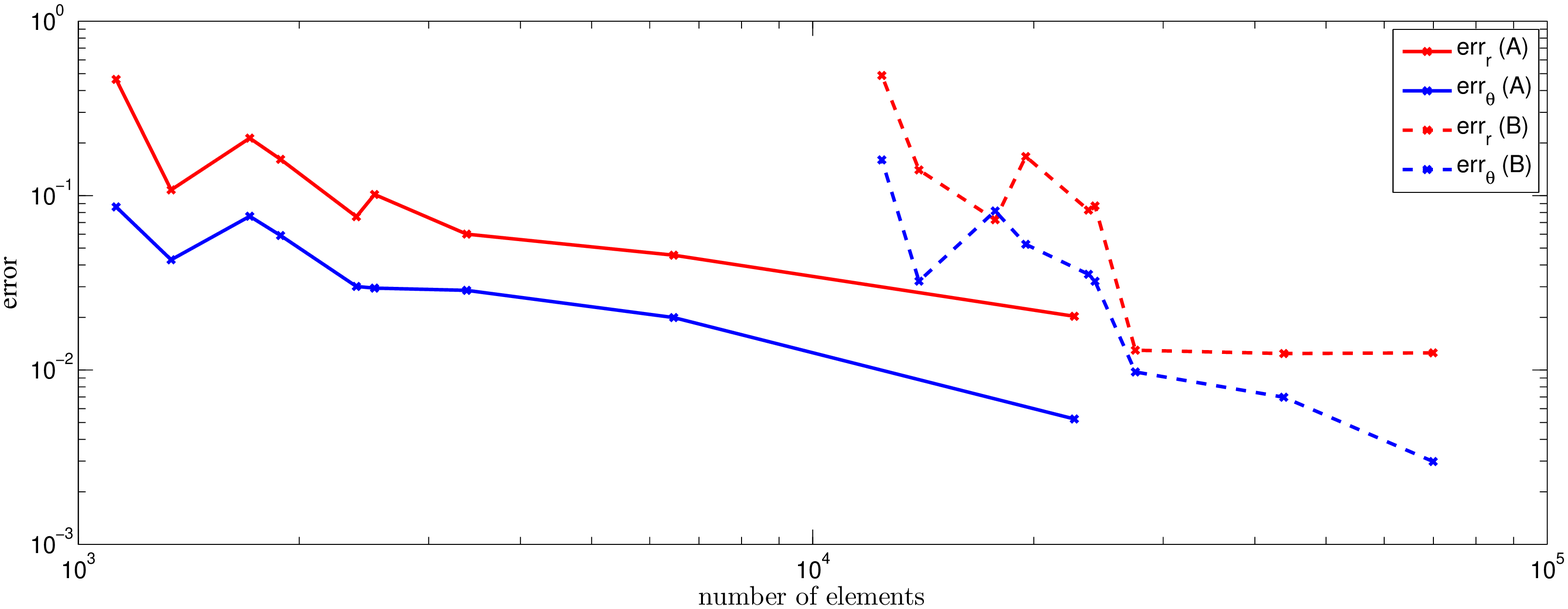}
\caption{Convergence of the FE simulation run through COMSOL Multiphysics\textsuperscript{\circledR} toward the analytical solutions. The case A shows the results for a thin layer of 20mm, the case B of 2mm placed in a full cylinder of radius 1m.}%
\label{fig:error42vs402}%
\end{figure}

The first benchmark computed via the present analytical exact solution is a multilayer structure of radius $R_2^{(n)}=1$m. The $i$-th interface is situated at $r=i{R_2^{(n)}}/{5}$ except for the third one which at $0.4\text{m}+t$. The thickness $t$ of the second layer is 2cm in the case A and is 2mm in the case B. The Young modulus and Poisson's ratio and the thermal load of the whole body are 30GPa, 0.222 and 907\textdegree K; the coefficient of thermal expansion is alternately $40\text{\textdegree}K^{-1}$ and $8\text{\textdegree}K^{-1}$ from the central component outwards. The plot in Fig.\ref{fig:reshoop5} shows the good agreement, for the case A, of the results both in terms of radial and hoop stresses if compared to a finite element (FE) simulation. For the case B such comparison would remain, for practical purposes, unvaried. The situation changes if one refers to the errors. The dimensionless error of the FE solution with respect to the analytical one (AN) for the stresses $\sigma_{ii}$ ($i=r,\theta$) is calculated as
\begin{equation*}
\text{err}_i=\frac{\sqrt{\sum{(\sigma_{ii}^{(\text{FE})}-\sigma_{ii}^{(\text{AN})})^2}}}{\max|\sigma_{ii}^{(\text{AN})}|}.
\label{eq:}
\end{equation*}
In Fig.\ref{fig:error42vs402} one can observe that, due to the tenfold reduced thickness of the layer $(2)$ in the case B, the mesh required in a FE analysis to approximate the exact solution at the same level of accuracy is much finer than in the case A. As evident consequence, a highly increased computational effort is required and special domain decomposition techniques may be required (e.g. \cite{dornisch2015weak}). This is, after all, the reason for the introduction of imperfect interface conditions in presence of very thin layers in a similar fashion as derived in \cite{avila1999asymptotic,benveniste1985effective,klarbring1998asymptotic,mishuris2004imperfect,benveniste2001imperfect}. The advantages of such mathematical approach have been widely exploited in engineering computations, demonstrating to be highly effective via finite and boundary elements analysis, by the way, applied to the design of elastic structures (\cite{mishuris2005fem,mishuris2006fem}), materials tests (\cite{ochsner2007new}) or to improve the numerical performances in thermo-diffusive contact problems (\cite{wrobel2009numerical,wrobel2009numericalb}).

The simple mechanical model explored in the Section\ref{sec:mechmod} proves to be fruitful in characterizing a multilayer cylindrical ceramic structure, particularly attaining to production precautions. The layers are obtained usually by overlapping different ceramic powders and sintering the previously compacted so-called \textit{green body} (e.g. \cite{rahaman2006ceramic}). The goal is to fuse the constituent grains together. A cooling process starts thereafter, which must be accurately controlled in order not to give place to destructive fractures due to unexpected self-stress (\cite{r14}).

Lugovy et al. (\cite{r8}) presented the concept of \textit{joining temperature} $T_{\text{join}}$: a temperature below which, during the cooling process, two phases start to deform together under an ideal contact condition. The materials difference in thermal expansion properties and stiffness are the main origin of residual stresses inside the final product and such mismatches evolve during the temperature reduction.
\begin{figure}%
\centering
\includegraphics[width=\textwidth]{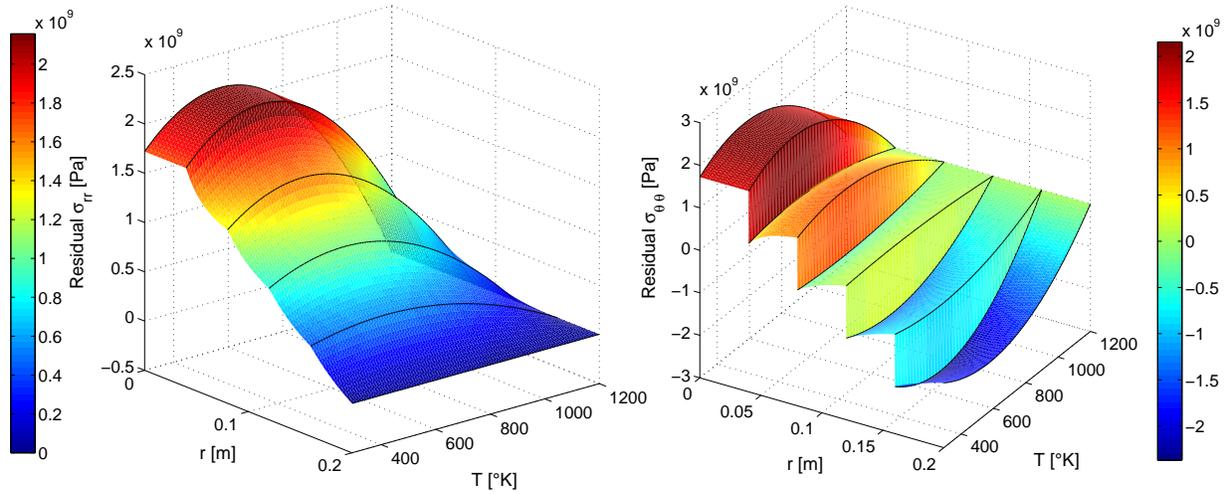}
\caption{Evolution of the residual stresses with decreasing temperature from $T_{\text{join}}$ to $T_{\text{room}}$. $E=E_{\text{ref}}(T)$ everywhere, $\hat{\alpha}^{(j)}=(5-j)\hat{\alpha}_{\text{ref}}(T)$ with $j=0:4$.}%
\label{fig:5elnosoft}%
\end{figure}
\comm{Taking advantage of the analytical tool derived in the previous section, we show hereafter the qualitative influence of the mechanical properties of the constituent layers for explaining different possible scenarios in terms or residual stresses. }For the computation, we assume that:
\begin{itemize}
	\item the thermal expansion coefficients increase with the temperature, both Young's moduli and the stress resistance vice versa, while Poisson's ratios remain constant with that respect (\cite{r7, r15});
	\item the above dependencies are linear;
	\item the cooling is performed relatively slowly so that the temperature gradient inside the structure at every moment is negligible. 
	\end{itemize}
It should be remarked that, in this analysis, the temperature has to be lower than the one at which the phase changes occur, since the positions of the interfaces must be known. It has been indeed proofed, based on thermodynamical considerations, that the combination of interface conditions  Eq.\eqref{eq:D1D2}-\eqref{eq:S1S2} or Eq.\eqref{eq:stD1D2}-\eqref{eq:stS1S2} and boundary conditions Eq.\eqref{eq:extBC} require to be treated accurately in the case of phase transition in order to ensure the stability of the solution (see \cite{yeremeyev2007stability}).

The geometry of the system for the following two benchmarks at the joining temperature is the same: a multilayer cylinder of radius $20$cm is composed of equally spanned layers.\comm{ The same tensile and compressive strength for each layer, depending on the temperature $T$[\textdegree K] is $\sigma_f[GPa]=39-0.004T$.} Poisson's ratio is $0.3$ everywhere. A joining temperature $T_{\text{join}}=1200.15$\textdegree K is considered for every interface. The reference temperature-dependent thermal expansion coefficient and Young's modulus are $\hat{\alpha}_{\text{ref}}(T)=10^{-6}(1.5+0.037T) \text{\textdegree K}^{-1}$ and $E_{\text{ref}}=(390-0.02T)$Gpa. The structure is cooled uniformly down to room temperature $T_{room}=293.15$\textdegree K and the differences in material properties give place to residual stresses during the process. Attention is payed to the surface stresses which can be responsible of toughening at the macroscale or of changing the elastic response of the material at nanoscale, as for instance discussed in \cite{altenbach2013influence,eremeyev2015effective}.
	
Five different materials are assembled whose thermal expansion coefficients decrease 5 times from the surface layer toward the inner cylinder while every other parameter is kept constant along the radius. Such an arrangement causes the temperature stresses upon cooling plotted in Fig.\ref{fig:5elnosoft}. It has been shown in \cite{r1, r13} that the presence of compressive residual stresses results advantageous for enhancing crack bifurcation if the compression is perpendicular to the crack growth. Considering a fracture which propagates from the external surface inwards, the presence of residual compressive hoop stresses in the superficial layers is able to increase the fracture toughness. The scope appears to be reached thanks to the special arrangement hypothesized here. On the other hand, one can observe that tensile radial stresses at the interfaces may raise the risk of delamination. Moreover, the shielding from radial crack growth is not guaranteed in the case the crack nucleates in the core of the structure, which may be the understanding key, for instance, of the disintegrating fracture of Prince Rupert's drops (\cite{r3,r6}).
		
\begin{figure}%
\centering
\includegraphics[width=\textwidth]{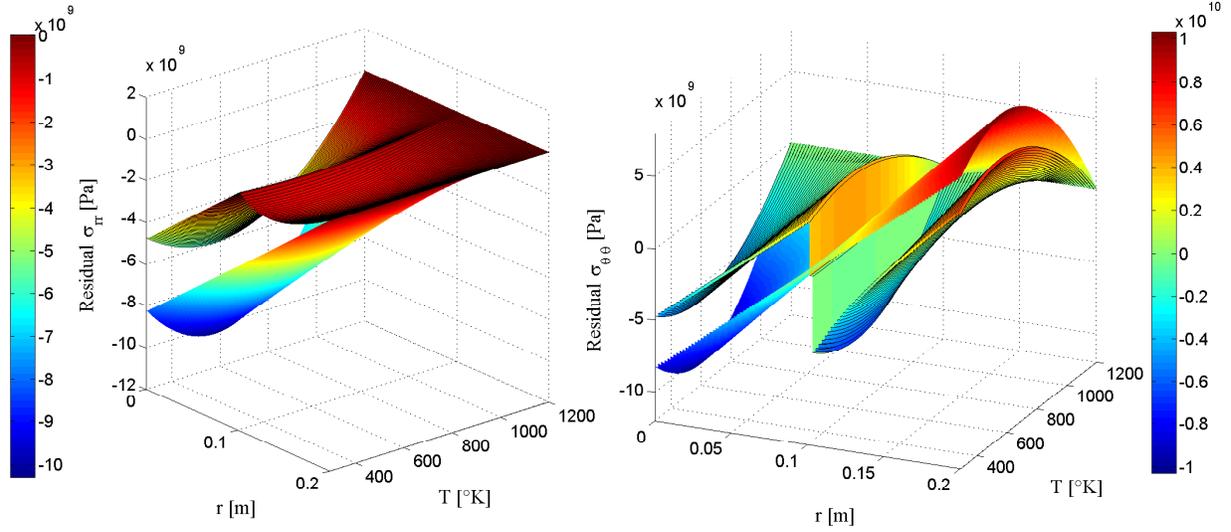}
\caption{Evolution of radial and hoop residual stresses with decreasing temperature from $T_{\text{join}}$ to $T_{\text{room}}$ for a graded material in terms of thermal expansion coefficients whose $\alpha$decreases 5 times from the outer surface to the symmetry axis. Comparison between homogeneous Young's modulus $E_{\text{ref}}$ (unshaded surfaces) and inhomogeneous with a soft internal layer of $E=10^{-3}E_{\text{ref}}$ (hatched surfaces).}%
\label{fig:100soft}%
\end{figure}

The structure for which the residual stresses are shown in Fig.\ref{fig:100soft} constituted of a graded material in the sense of thermal expansion coefficient. It ranges linearly along the radius from $\hat{\alpha}_{\text{ref}}$ at the symmetry axis to $5\hat{\alpha}_{\text{ref}}$ at the external surface -- thus on the contrary as the example of Fig.\ref{fig:5elnosoft} -- and discretized into 100 layers.
As explained in the Section \ref{sec:intro}, a common artifice to arrest the crack propagation is the provision of weak interlayers inside a brittle structure. The materials used to this purpose for engineering application or found in nature  (e.g. see \cite{munch2008tough}) often present lower stiffness than the surrounding ones and larger failure strain with the result of an increased fracture work. In this benchmark, one of the layers is a soft thin circular layer introduced in the middle of the overall radius and it is characterized by a value of Young's modulus of $10^{-3}E_{\text{ref}}$, where $E_{\text{ref}}$ is assigned to each other component. The behavior of such structure upon cooling, pictured in hatched surfaces in Fig.\ref{fig:100soft}, is compared to the same structure in absence of the soft layer, plotted in unshaded surface in the same figure. The effect of the soft element is a jump in the hoop stresses which allows to reduce, as expected, the surface tension of the cylinder, as well as approaching a compressive zone to the same surface which may work, as discussed dealing with the previous benchmark, to arrest a radial crack. However, it must be noticed that the residual radial compression is weakened, which is crucial when protecting the structure from delamination. It is the engineer who must choose, in the design phase, to what kind of danger risk the product will be most probably subjected during its service life and adopt an appropriate safety criterion, pragmatically via an optimization process.
	
\section{Conclusions}
The plain strain deformation problem for a multilayer cylinder composed of an inner cylinder inscribed by a general number of annular layers is solved analytically in the case of mechanical loads or constraint and thermal expansion. The proposed solution includes also prestresses and three kinds of interface transmission conditions: perfect, membrane-type and spring-type. The reasons for the introduction of the latter feature are made clear through a study of the disadvantages of a FE solution in presence of thin layers due to important mesh refinement required in the vicinity of the interfaces (see Fig.\ref{fig:error42vs402}). In that case, indeed, the computational efforts for reaching an acceptable accuracy largely increase with the reduced ratio between the thin layer size and the adjacent ones. The set of linear equations deriving from the present formulation has been solved efficiently applying a sweep method and illustrated in \ref{sec:swm}. 
The method was adopted for studying the uniform cooling of multilayer cylinders whose composing materials starts to deform together below their joining temperature and whose thermoelastic parameters are temperature-dependent. Based on literature considerations regarding fracture toughness of composites, mainly ceramics, selected numerical examples show how different arrangements of thermal expansion coefficients or the inclusion of soft layers inside a stiffer environment are able to create residual stresses which can shield the structure from crack propagation. 

\section*{Acknowledgments}
The Authors are grateful to the European Union’s Seventh Framework Programme FP7/2007-2013/, REA grant agreement number PITN-GA-2013-606878-CERMAT2, under whose financial support the present work was accomplished.

\section*{References}
\bibliographystyle{abbrv}

\bibliography{mybibfile}

\appendix
\section{A Sweep Method}
\label{sec:swm}
Based on the approach adopted in \cite{linkov1992difference} for layered systems and expanded in \cite{mishuris19972} to wedged-layered domains, for the sake of computational efficiency, we propose here a sweep method to solve the linear system stated in Eq.\eqref{eq:system}. Condensing the writing, we set here the following abbreviations:
\begin{equation}\left\{
\begin{aligned}
&\VE{M}_i=\VE{M}_2^{(i)};\\
&\VE{G}_i=-\VE{M}_1^{(i)};\\
&\VE{B}=\VE{B}_2^{(n)};\\
&a_0=a^{(0)};  &\VE{a}_i=\left[\begin{array}{c}a^{(i)}\\b^{(i)}\end{array}\right];\\
&d_e=d_{2n+1};&\VE{d}_i=\left[\begin{array}{c}d_{2i-1}\\d_{2i}\end{array}\right];
\end{aligned}
\right.\end{equation}
Therefore, the system can be expressed as
\begin{equation}
\left[\begin{matrix} \VE{M}_0&\VE{G}_1\\
&\VE{M}_1&\VE{G}_2\\
&&\ddots&\ddots\\
&&&\VE{M}_{n-1}&\VE{G}_n\\
&&&&\VE{B}
\end{matrix}\right]
\left[\begin{matrix} a_0\\ \VE{a}_1\\ \VE{a}_2\\ \vdots \\ \VE{a}_n 
\end{matrix}\right]=
\left[\begin{matrix} \VE{d}_1\\ \VE{d}_2\\ \vdots \\ \VE{d}_n \\ d_e 
\end{matrix}\right].
\end{equation}
It is a bidiagonal block matrix. $\VE{B}\in\mathbb{R}^{1\times 2} $ and $\VE{M}_0\in\mathbb{R}^{2\times 1}$ are not invertible, thus the method shown here is devised to fix this features.
As start of forward sweeping, the second matrix equation is substituted by $\VE{G}_1 \VE{M}_1^{-1}$ times itself minus the first equation. The variable $\VE{a}_1$ is eliminated, the variable $a_0$ appears:
\begin{equation}
-\VE{M}_0 a_0+\VE{G}_1 \VE{M}_1^{-1} \VE{G}_2 \VE{a}_2=\VE{G}_1 \VE{M}_1^{-1}\VE{d}_2-\VE{d}_1 .\\
\end{equation}
Or, defining $\VE{G}'_2=\VE{G}_1 \VE{M}_1^{-1} \VE{G}_2$ and $\VE{d}'_2=\VE{G}_1 \VE{M}_1^{-1}\VE{d}_2-\VE{d}_1$:
\begin{equation}
-\VE{M}_0 a_0+\VE{G}'_2 \VE{a}_2=\VE{d}'_2 .\\ 
\end{equation}
The method is substantially based on vanishing $\VE{a}_{i-1}$ and including $a_0$ in the $i$-th equation ($i=2,...,n$). The forward sweep, posed that $\VE{G}'_1=\VE{G}_1$ and $\VE{d}'_1=\VE{d}_1$, is therefore described by the following recursive formulae:
\begin{equation}
\text{for }i=2,...,n\quad
\left\{
\begin{aligned}
&\VE{G}'_i={\VE{G}'}_{i-1}{\VE{M}_{i-1}}^{-1}\VE{G}_i\\
&\VE{d}'_i={\VE{G}'}_{i-1}{\VE{M}_{i-1}}^{-1}\VE{d}_i-\VE{d}'_{i-1}
&\end{aligned}\right. ,
\end{equation}
so that the $i$-th equation takes the form
\begin{equation}
(-1)^n\VE{M}_0 a_0 + \VE{G}'_i \VE{a}_{i}=\VE{d}'_i .
\label{eq:sweptg}
\end{equation}
The $n$-th Eq.\eqref{eq:sweptg}, combined with the outer boundary condition, gives the $3\times 3$ linear set of equations
\begin{equation}
\left\{
\begin{aligned}
&(-1)^i\VE{M}_0 a_0 + \VE{G}'_n \VE{a}_{n}=\VE{d}'_n\\
&\VE{B}\VE{a}_n=d_e
&\end{aligned}\right. ,
\end{equation}
whose solution is
\begin{equation}
\left[\begin{matrix}a_0\\ \VE{a}_n \end{matrix}\right]=
\left[\begin{matrix}(-1)^n\VE{M}_0 & \VE{G}'_n\\ 0&\VE{B} \end{matrix}\right]^{-1}
\left[\begin{matrix}\VE{d}'_n\\ d_e \end{matrix}\right].
\end{equation}
With $\VE{a}_n$ in hand, a backward sweep can be started which finds the remaining $n-1$ vector variables
\begin{equation}
\VE{a}_{i}=\VE{M}_{i}^{-1}(\VE{d}_{i+1}-\VE{G}_{i+1}\VE{a}_{i+1}) ,\\
\end{equation}
for $i=n-1,...,1$. It is probably worth to mention that the matrices $\VE{M}_i^{-1}$ are already computed during the forward sweep.

\end{document}